\documentclass{new-aiaa}
\DeclareUnicodeCharacter{202F}{\,}
\usepackage[utf8]{inputenc}

\usepackage{graphicx}
\usepackage{amsmath}
\usepackage[version=4]{mhchem}
\usepackage{hyperref}
\usepackage{siunitx}
\usepackage{comment}
\usepackage{longtable,tabularx}
\usepackage{float}
\usepackage{booktabs}
\usepackage{multicol}
\usepackage{multirow}
\usepackage{enumitem}
\usepackage{array}
\usepackage{makecell}

\title{A Survey of Security Challenges and Solutions for \\ UAS Traffic Management (UTM) and \\ small Unmanned Aerial Systems (sUAS)}

\author{Iman Sharifi, Mahyar Ghazanfari, Abenezer Taye, Peng Wei}
\affil{The George Washington University, Washington, DC, 20052, USA}
\author{Maheed H. Ahmed, Hyeong Tae Kim, Mahsa Ghasemi, Vijay Gupta}
\affil{Purdue University, West Lafayette, IN, 47907, USA}
\author{Noah Dahle, Robert Canady, Abel Diaz Gonzalez, Austin Coursey, Bryce Bjorkman, Cailani \\ Lemieux-Mack, Bryan C. Ward, Xenofon Koutsoukos, Gautam Biswas}
\affil{Vanderbilt University, Nashville, TN, 37212, USA}
\author{Heber Herencia-Zapana, Saqib Hasan, Isaac Amundson}
\affil{Advanced Research \& Technology, Collins Aerospace, Cedar Rapids, IA, 52498, USA}
\author{Filippos Fotiadis, Ufuk Topcu}
\affil{The University of Texas at Austin, Austin, TX, 78712, USA}
\author{Junchi Lu, Qi Alfred Chen}
\affil{University of California, Irvine, Irvine, CA, 92697, USA}
\author{Nischal Aryal, Amer Ibrahim, Abdul Karim Ras, Amir Shirkhodaie}
\affil{Tennessee State University, Nashville, TN, 37221, USA} 

\begin{document}

\maketitle

\begin{abstract}
The rapid growth of small Unmanned Aerial Systems (sUAS) for civil and commercial missions has intensified concerns about their resilience to cyber–security threats. Operating within the emerging UAS Traffic Management (UTM) framework, these lightweight and highly networked platforms depend on secure communication, navigation, and surveillance (CNS) subsystems that are vulnerable to spoofing, jamming, hijacking, and data manipulation. While prior reviews of UAS security addressed these challenges at a conceptual level, a detailed, system-oriented analysis for resource-constrained sUAS remains lacking. This paper presents a comprehensive survey of cyber-security vulnerabilities and defenses tailored to the sUAS and UTM ecosystem. We organize existing research across the full cyber–physical stack, encompassing CNS, data links, sensing and perception, UTM cloud access, software integrity layers, and classify attack vectors according to their technical targets and operational impacts. Correspondingly, we review defense mechanisms ranging from classical encryption and authentication to adaptive intrusion detection, lightweight cryptography, and secure firmware management. By mapping threats to mitigation strategies and evaluating their scalability and practical effectiveness, this work establishes a unified taxonomy and identifies open challenges for achieving safe, secure, and scalable sUAS operations within future UTM environments.
\end{abstract}

\section{Introduction}

\smallskip 

\textit{Unmanned Aerial Systems (UASs)} are increasingly being adopted across a wide range of commercial and industrial domains, including infrastructure inspection, aerial surveying, environmental monitoring, and package delivery \cite{aerospace8120363}. The proliferation of low-altitude UAS operations presents both promising opportunities and significant challenges for integrating these systems into the existing regulatory and operational framework of the National Airspace System (NAS). In response to the growing demand and complexity of UAS operations, particularly those conducted at scale and beyond visual line of sight (BVLOS), the \textit{Federal Aviation Administration (FAA)} and the \textit{National Aeronautics and Space Administration (NASA)} are leading a joint initiative to develop a comprehensive \textit{UAS Traffic Management (UTM)} framework \cite{kopardekar2016unmanned}. UTM is envisioned as a scalable, service-based architecture designed to safely accommodate the increasing volume and diversity of UAS operations without overburdening the existing Air Traffic Management (ATM) infrastructure \cite{faautm2020}. Within this evolving ecosystem, small Unmanned Aerial Systems (sUAS) \cite{https://doi.org/10.1111/j.1937-2817.2010.tb01292.x} represent a rapidly expanding class of vehicles, with weights under 55lbs, whose compact design, low cost, and operational flexibility make them particularly well-suited for high-frequency, localized missions \cite{tang2021review}. 

Moreover, the rapid evolution of drone technology has resulted in a surge of interest in sUAS applications across various sectors. Companies like Google Wing \cite{zenz2024resisting}, Amazon Prime Air \cite{jung2017analysis}, Zipline \cite{ackerman2019blood}, and UPS Flight Forward are actively deploying or testing drone-based delivery systems for pharmaceuticals, groceries, and e-commerce goods. For instance, Zipline's fixed-wing drones have revolutionized medical supply chains in countries like Rwanda and Ghana by reliably delivering blood and vaccines over long distances \cite{ackerman2019blood}. These advancements demonstrate that sUAS are no longer experimental platforms but viable elements of real-world air logistics. However, as these systems begin to share increasingly complex and populated airspace under the UTM framework, safety becomes an indispensable requirement—not just in terms of physical collision avoidance, but also in securing digital communication and control channels \cite{icao2018utm}. This transition introduces new concerns related to cyber-security, which must be addressed to maintain trust, reliability, and regulatory compliance in UTM operations~\cite{nasa2021maap}.

As sUAS operations expand in both scale and complexity, their increasing dependence on digital connectivity exposes them to a diverse range of cyber–security threats~\cite{nasa2024dss}. Because these aircraft rely on telemetry link, command and control (C2) links, GPS-based navigation, remote ID~\cite{faa_remote_id}, and cloud-enabled UTM services, their operational integrity depends on the security of underlying communication, navigation, and surveillance subsystems \cite{8396511}. Compromises in any of these layers can cascade into loss of control or mission failure. Incidents of GPS spoofing, Radio Frequency (RF) jamming, and denial-of-service attacks have demonstrated that even short-term disruptions in connectivity or sensor integrity can endanger flight safety and compromise public trust. The unique size, weight, and power (SWaP) constraints of sUAS exacerbate this problem by limiting the adoption of conventional cyber-security solutions designed for larger aerial or terrestrial systems.

Previous survey efforts have examined cyber-security concerns within broader contexts of aerial mobility and autonomous flight. Among them, Tang~\textit{et al.}~\cite{tang2021review} conducted a systematic review of cyber-security vulnerabilities relevant to Urban Air Mobility (UAM), consolidating known attacks across UAS and commercial aviation systems. While this work provided an important foundation by highlighting the necessity of secure data links and resilient architectures, its discussion remained primarily conceptual and focused on large passenger or cargo-carrying vehicles operating in controlled airspace. In contrast, the cyber-security landscape of sUASs introduces a distinct set of challenges arising from constrained hardware resources, heterogeneous platforms, decentralized UTM coordination, and frequent low-altitude operations in dynamic environments. These factors demand a more granular and technically grounded examination of attack surfaces, communication dependencies, and defense mechanisms specific to sUAS networks.

To address this gap, this paper extends the scope and analytical depth of prior studies by providing a detailed, system-oriented review of cyber-security vulnerabilities and countermeasures within the sUAS and UTM ecosystem. Our study differs from earlier UAM-focused surveys by explicitly mapping cyber threats to the underlying communication, navigation, and surveillance (CNS) subsystems that govern sUAS operations. We further incorporate perception and software integrity as additional layers of analysis, capturing the full cyber–physical interaction chain from sensing and decision-making to remote command execution. For each subsystem, we compile and classify attack vectors reported in academic literature, incident databases, and industrial test reports, linking them to specific operational consequences such as data corruption, control loss, and mission interruption. In parallel, we review both conventional defenses—such as encryption, authentication, and spread-spectrum communication—and emerging techniques, including adaptive intrusion detection, lightweight cryptography, and secure update frameworks. By organizing this material into a unified taxonomy of vulnerabilities and mitigations, our survey aims to (1) identify recurring attack patterns across heterogeneous sUAS architectures, (2) evaluate the effectiveness and scalability of existing countermeasures, and (3) outline open challenges for future secure integration under the evolving UTM framework.

The remainder of this paper follows this organization. Section~II introduces the baseline architectures and data flows that define sUAS and UTM operations. Section~III provides a taxonomy of vulnerabilities and corresponding real-world incidents across CNS, perception, and software domains and surveys existing defense mechanisms and identifies persistent gaps and research opportunities. Finally, Section~IV draws a conclusion for secure and scalable integration of sUAS into future UTM ecosystems.

\begin{figure}[ht]
    \centering
    \includegraphics[width=0.7\linewidth, clip, trim=3mm 3mmm 3mm 5mm]{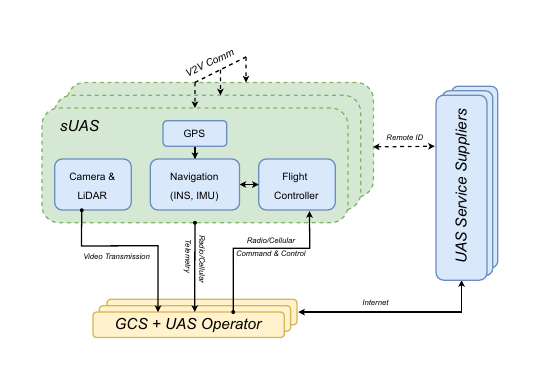}
    \caption{Representative architecture of sUAS, ground control station (GCS) and UAS service suppliers (USS). The solid lines represent existing data links and connections. The dashed lines represent envisioned data links and connections.}
    \label{fig:suas_architecture}
\end{figure}

\section{Background: Architectures of sUAS and UTM}
\smallskip

sUASs operate through an interconnected set of hardware and software components that enable autonomous or automated flight in low-altitude airspaces. These systems are supported around three critical functional areas: communication, navigation, and surveillance (CNS), on top of the aircraft autonomy stack. Together, the CNS functions facilitate the transmission of information, the localization of the aircraft, and the visibility of operations to external entities such as traffic service providers or other operators.

\textbf{Communication} in sUASs is the cornerstone of remote operation. It comprises the command and control (C2) links as well as telemetry and video transmission systems. The C2 link is typically established through radio frequencies, commonly at 2.4 GHz or 5.8 GHz, and serves as the primary medium through which operators issue commands to the aircraft. In more advanced systems, cellular connections such as 4G or 5G may be employed to enable beyond visual line of sight (BVLOS) operations. Telemetry and video data are often transmitted through the same RF bands, either on separate channels or via a combined data link, depending on the architecture. A typical sUAS setup involves a remote controller or ground control station (GCS), a wireless transceiver onboard the aircraft, and sometimes an onboard flight computer that aggregates sensor data, manages autonomous flight logic, and relays information back to the operator.

\textbf{Navigation} is largely handled through the Global Positioning System (GPS), which provides real-time location and velocity data to the aircraft. Most sUASs use GPS as the sole navigation source for tasks such as waypoint tracking, geofencing, and return-to-home (RTH) protocols. Some UASs also incorporate inertial navigation systems (INS) that provide positional data through onboard sensors like gyroscopes and accelerometers (e.g. IMU). These redundant systems are critical in the event of GPS signal loss or spoofing. GPS signals, particularly civilian-grade signals, are unencrypted and thus susceptible to manipulation, making navigation a significant point of vulnerability.

\textbf{Surveillance}, the third pillar of CNS, involves the external monitoring of aircraft by ground-based systems. Unlike commercial aircraft, which are equipped with ADS-B transmitters for real-time broadcasting of location and identity, sUASs typically operate under different constraints. The use of ADS-B is generally discouraged or even prohibited for sUASs due to spectrum congestion and scalability limitations. Instead, the FAA mandates the use of Remote ID technology~\cite{faa_remote_id}. Under this framework, the aircraft periodically broadcasts its identity, position, velocity, and altitude. This information is collected by the GCS and expected to be uploaded to cloud-based UTM platforms or FAA databases for monitoring. UAS Service Suppliers (USSs) would play an essential role in this process by aggregating and exchanging Remote ID and airspace information to ensure cooperative surveillance across the UTM ecosystem. The Remote ID system is essential for integrating sUASs into the national airspace without overloading existing air traffic surveillance infrastructure.

The internal architecture of an sUAS typically includes an onboard computer that processes data and governs flight logic, a flight controller helps track the planned trajectory and maintain flight stability, a GPS module for positioning, a suite of sensors such as Inertial Measurement Unit (IMU), cameras and LiDAR, radio or cellular transceivers as well as remote ID transmitters, that facilitate data exchange with the ground station. These components are interconnected via internal wiring, while external communication occurs over RF or cellular links. Figure~\ref{fig:suas_architecture} illustrates a typical architecture for sUASs, GCSs and their USSs, highlighting the various information/data links and subsystem modules.


The dependence of sUASs on unsecured RF links and unencrypted GPS signals makes them inherently vulnerable to jamming, spoofing, and signal interception. GPS spoofing can mislead the aircraft about its true location, while RF jamming can disturb the C2 link, potentially grounding the aircraft or causing it to fly unpredictably. Man-in-the-middle (MitM) attacks are also feasible, wherein an adversary intercepts and modifies commands or telemetry in real time. Remote ID is highly unsecured. These vulnerabilities are further compounded by the diversity of manufacturers and the lack of standardization in securing these systems. Many consumer drones use proprietary methods for linking remote controllers to aircraft, and video feeds are often transmitted without encryption.

As the number of sUAS operations increases and their integration into the low-altitude airspaces progresses, ensuring the security of CNS and autonomy functions becomes critical. Without robust safeguards, these aircraft could be exploited to compromise privacy, disrupt services, or cause physical harm. Therefore, understanding and fortifying the CNS and autonomy architectures of sUASs is a foundational step toward building a secure, scalable, and resilient unmanned traffic management (UTM) ecosystem.



\section{Existing Vulnerabilities and Defense Mechanisms}
\label{vulnerabilties}
\smallskip 

Similar to commercial aviation, sUASs and UTM systems are highly susceptible to various forms of cyber threats \cite{aviation_2022, aviation, spicejet_2022, canadian_contractor, boeing_subsidiary_2022, aviation_cs, aviation_2023}. These cyber threats exploit vulnerabilities that are present in various parts of the sUASs and UTM systems individually. In order to investigate sUAS-associated cyber-security vulnerabilities and the corresponding defense mechanisms, we consider all security subsystems of an sUAS for comprehensive system-level security analysis. As shown in Figure \ref{fig:suas-security-exploitations}, the sUAS subsystems that have security vulnerabilities include: communication, navigation, perception, and software subsystems. Each subsystem can be compromised through various attacking strategies, and at the same time there are preventive defense mechanisms to mitigate the existing threats. For each subsystem, we categorize the existing threats and for each threat, we present the current defense mechanisms as well as the research challenges and open problems. Below are the major categories of the subsystem vulnerabilities that are worth considering:

\begin{figure}
    \centering
    \includegraphics[width=\linewidth]{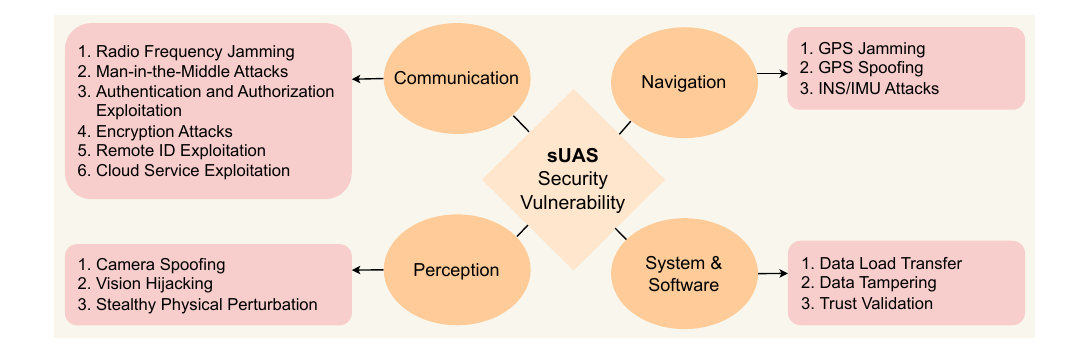}
    \caption{sUAS security exploitations from four different perspectives: Communication, Navigation, Perception, as well as System- and Software-level vulnerability exploitation.}
    \label{fig:suas-security-exploitations}
\end{figure}

\subsection{\textit{Communication Vulnerabilities and Defense Mechanisms}}

This section focuses on all vulnerabilities targeting the communication infrastructure of sUASs and UTM systems, the relevant defense mechanisms, and the current gaps in the literature. It includes both wireless communication links (e.g., C2, telemetry, Remote ID) and network-based data exchanges (e.g., over-the-air updates, GCS connections to USS). Detailed explanations of communication vulnerabilities and corresponding defense mechanisms with open problems are discussed in the following subsections: 

\subsubsection{\textbf{Radio Frequency Jamming}}


\textbf{\textit{Existing Threats:}} Radio-frequency (RF) jamming refers to the intentional emission of interference to disrupt wireless communication channels. Classical taxonomies categorize jammers into several operational modes: \emph{constant (barrage) jammers}, which continuously flood the channel with wideband noise; \emph{reactive jammers}, which transmit only when legitimate activity is detected; and \emph{deceptive or protocol-aware jammers}, which craft signals that mimic valid frames or control channels~\cite{lichtman2016taxonomy,pelechrinis2011jammers}. These behaviors can be instantiated as multi-tone, sweep, or adaptive “smart” jammers capable of tracking physical-layer features of the victim link.

sUASs rely heavily on unlicensed 2.4\,GHz and 5\,GHz bands for C2 and telemetry, making them especially vulnerable to opportunistic or targeted interference. Because these links share spectrum with Wi-Fi and other consumer devices, even moderate-power jammers can induce high packet-error rates or complete link loss~\cite{pelechrinis2011jammers}. Cellular-based C2 links also remain susceptible to uplink jamming in the vicinity of the UAS.

Loss of C2 link can trigger emergency behavior such as hover, return-to-home, or forced landing, depending on autopilot configuration. Global Navigation Satellite System (GNSS) disruption—through jamming or spoofing—can further degrade positioning accuracy, as demonstrated experimentally by Tippenhauer \textit{et al.}~\cite{tippenhauer2011gpsspoofing} and through stealthy trajectory-manipulation strategies in~\cite{su2016stealthygps}. Broader surveys confirm that GNSS interference can lead to navigation drift, geofence violations, and mission aborts~\cite{dulowicz2025uavgnss}.

Attackers can exploit a range of readily available tools, including low-cost software-defined radios (SDRs), wideband noise generators, sweep jammers, or deep-learning–supported deception systems~\cite{simon2022uavdl}. Real-world demonstrations show that commercial multicopters can lose C2 or be forced into failsafe modes under modest interference levels~\cite{simon2022commercialuav}. Table~\ref{table:jamming-types} summarizes different jamming types with the affected systems as well as the operational consequences after jamming.

\begin{table}[ht]
\centering
\caption{Summary of Jamming Types, Affected Systems, and Operational Consequences}
\begin{tabular}{p{3cm} p{4cm} p{8cm}}
\hline
\textbf{Jamming Type} & \textbf{Target Systems} & \textbf{Main Consequences} \\
\hline
Constant/Barrage & C2, Telemetry, GNSS & Continuous denial; failsafe activation; high jammer power~\cite{lichtman2016taxonomy}. \\
Reactive & C2 bursts & High PER with low duty cycle; harder to detect~\cite{lichtman2016taxonomy}. \\
Deceptive/Protocol-aware & C2 control/data channels & Forged frames, desynchronization, takeover attempts~\cite{pelechrinis2011jammers}. \\
Sweep/Tone & FHSS or multi-channel systems & Intermittent disruption; frequent channel reacquisition~\cite{lichtman2016taxonomy}. \\
Adaptive/Smart & FHSS, adaptive radios & Low-power, high-efficiency interference; can track patterns~\cite{simon2022uavdl}. \\
\hline
\end{tabular}
\label{table:jamming-types}
\end{table}



\textbf{\textit{Defense Mechanisms:}} Anti-jamming defenses for sUAS span traditional spread-spectrum methods, adaptive resource-allocation mechanisms, and emerging machine-learning techniques. \emph{Frequency-Hopping Spread Spectrum (FHSS)} mitigates narrowband interference by rapidly switching carrier frequencies according to a shared sequence. \emph{Direct-Sequence Spread Spectrum (DSSS)} spreads each symbol with a pseudo-random code, lowering jammer efficiency by increasing processing gain~\cite{pelechrinis2011jammers}. These techniques remain attractive for sUAS due to their low computational overhead and minimal hardware burden.

Beyond spread spectrum, \emph{adaptive power control} can temporarily improve link reliability but is constrained by sUAS battery capacity. \emph{Beamforming} and \emph{multi-input-multi-output (MIMO)} approaches provide spatial nulling to attenuate interference, though the antenna arrays and RF chains required are often incompatible with the SWaP limitations of small multirotors.

More advanced defenses borrow from game theory and reinforcement learning. Xu \textit{et al.}\ show that UASs operating on single-channel links can improve resilience by adopting game-theoretic anti-jamming strategies~\cite{xu2021antijamming}. Similarly, reinforcement-learning–based allocation schemes have demonstrated improved throughput and robustness in UAS relay networks subjected to interference~\cite{wu2019rlantijam}. Recent surveys also highlight the promise of \emph{machine-learning–based jamming detection}, which can identify interference signatures from RF or protocol-layer features~\cite{simon2022uavdl,yang2025agentbased}.

\begin{table}[ht]
\centering
\caption{Anti-Jamming Techniques and Applicability to sUAS}
\begin{tabular}{p{4cm} p{5cm} p{6cm}}
\hline
\textbf{Technique} & \textbf{Principle} & \textbf{sUAS Suitability} \\
\hline
FHSS / DSSS & Spread signal across time/frequency/code & Highly suitable; lightweight and widely used~\cite{pelechrinis2011jammers}. \\
Adaptive Power Control & Boost TX power under interference & Limited by battery and PA constraints. \\
Beamforming / MIMO & Spatial nulling of jammers & Large SWaP footprint; unsuitable for small drones. \\
Multi-Link / Relaying & Provide redundant communication paths & Effective in networked or fleet operations~\cite{wu2019rlantijam}. \\
Game-Theoretic Approaches & Optimize strategy vs. adversarial jammer & Demonstrated effective in UAV C2 models~\cite{xu2021antijamming}. \\
ML-Based Detection & Classify jamming/spoofing via learned features & Emerging technique; promising but compute-limited~\cite{simon2022uavdl}. \\
\hline
\end{tabular}
\label{table:defenses}
\end{table}



\textbf{\textit{Defense Gaps and Open Problems:}} Existing anti-jamming strategies for sUAS face several limitations. Hardware-based defenses such as MIMO arrays or high-power amplifiers exceed SWaP constraints for sUASs. Spread-spectrum methods offer baseline resilience but can be circumvented by adaptive or protocol-aware jammers capable of sensing and reacting to hopping patterns or MAC-layer timing~\cite{lichtman2016taxonomy}. 

A significant challenge is the absence of standardized benchmarks for evaluating sUAS anti-jamming performance. Studies vary widely in jammer models, channel assumptions, and evaluation metrics, making cross-comparison difficult. The rapid evolution of intelligent jammers—including deep-learning–driven spoofers and cognitive interferers—further challenges static or handcrafted defenses~\cite{simon2022uavdl}.

Scalability also remains underexplored: dense UTM environments require cooperative spectrum sharing, distributed sensing, and fleet-level anti-jamming coordination. Emerging surveys emphasize the need for lightweight ML-based detectors, cross-UAS collaboration, and certifiable, low-SWaP anti-jamming protocols designed specifically for sUAS~\cite{yang2025agentbased,dulowicz2025uavgnss}.

Future research directions include: (1) cooperative anti-jamming across sUAS fleets; (2) hybrid GNSS–inertial spoofing detection suitable for low-cost autopilots; (3) lightweight onboard ML tailored to micro-UAS processors; and (4) standardized testbeds and metrics enabling reproducible evaluation across jammer types and operational settings.

\subsubsection{\textbf{Man-in-the-Middle (MitM) and Interception Attacks}} 


\textbf{\textit{Existing Threats:}}
The MitM attack is defined in the Common Attack Pattern Enumeration and Classification (CAPEC) \cite{mitreCAPEC}, as an attack that targets communication between two components—typically a client and a server—with the objective of intercepting or modifying transmitted data. Although this definition is general, it is directly applicable to sUASs, which often rely on unsecured or weakly protected communication links. In \cite{rodday2016exploring}, Rodday \textit{et al.} demonstrated that adversaries could exploit vulnerabilities in sUAS communication channels to perform MitM attacks and inject malicious control commands. Specifically, the study identified a weakness in the WiFi communication link, where the access point employed Wired Equivalent Privacy (WEP) encryption, a scheme known to be easily compromised. Several other studies \cite{yu2025electronic, koubaa2019micro} have also analyzed MitM attacks in sUAS environments, detailing associated vulnerabilities and confirming their practical feasibility.
Furthermore, CAPEC provides insights into the logical relationships between different attack types. For example, CAPEC-151 (Identity Spoofing) is often a subsequent step following a MitM attack (CAPEC-94). Identity Spoofing involves impersonating another entity—either human or non-human—to achieve a specific objective \cite{mitreCAPEC}. In the sUAS domain, Chamola \textit{et al.}~\cite{chamola2021comprehensive} reported that modern GPS receivers implement Automatic Gain Control (AGC) to compensate for signal fluctuations; however, this feature may inadvertently increase vulnerability to spoofing attacks. The observed relationship between MitM and subsequent spoofing illustrates how sequential attack chains can develop within sUAS communication systems.


\textbf{\textit{Defense Mechanisms:}}
From a defense perspective, several studies have proposed mitigation strategies. For example, Rodday \textit{et al.}~\cite{rodday2016exploring} recommend strengthening communication security by implementing robust encryption protocols for the WiFi 802.11 access point connecting the tablet to the telemetry unit and ensuring that data transmitted through the XBee 868LP chip is encrypted end-to-end. Similarly, Keshavarz \textit{et al.}~\cite{keshavarz2020real} propose a trust monitoring mechanism designed to detect MitM and other cyberattacks in real time. In their approach, the ground station continuously evaluates sUAS behaviors such as trajectory, energy consumption, and task performance to calculate dynamic trust scores and identify anomalies indicative of potential compromise. Simulation results demonstrate that this model can effectively detect malicious sUASs subjected to various cyber-security threats, including flooding attacks. Finally, as highlighted by \cite{gambo2025zero,alquwayzani2024systematic,ouiazzane2023zero}, the increasing complexity of digital ecosystems and the evolving nature of cyber-security threats expose the limitations of traditional perimeter-based security models. These challenges have driven the adoption of Zero Trust Architecture (ZTA) as a comprehensive defense paradigm that integrates encryption, authentication, continuous monitoring, and adaptive trust evaluation to secure sUAS networks and related cyber-physical systems.



\textbf{\textit{Defense Gaps and Open Problems:}} Several well-established databases provide structured information on threats, vulnerabilities, and attack patterns, including the MITRE Common Vulnerabilities and Exposures (CVE®), the NIST National Vulnerability Database (NVD), the MITRE Common Weakness Enumeration (CWE™), and the Common Attack Pattern Enumeration and Classification (CAPEC™). In terms of defense mechanisms, the Zero Trust (ZT) methodology has gained significant traction. ZT represents an evolving set of cyber-security principles that shift focus from traditional, perimeter-based defenses to user, asset, and resource-centric protection. One notable advantage of this approach is its well-defined taxonomy and terminology for attacks, vulnerabilities, and defenses. These standardized definitions establish clear relationships among these elements. However, in the sUAS domain, despite the use of such taxonomies for analysis, attacks are still rarely reported in existing repositories \cite{mitre2023uas_security}. Establishing a unified sUAS-specific database would therefore provide substantial benefits, enabling researchers to perform systematic analysis and cross-study comparisons. The study by MITRE \cite{mitre2023uas_security} provides a comprehensive overview of sUAS security, examining threats, vulnerabilities, and defenses while referencing these key databases. The authors note that sUAS-specific attacks are often not reported in such repositories, emphasizing the need for a unified global database to facilitate consistent research and comparative analysis. With this, some researches such as \cite{papoutsakis2025sesame}, employ CWE to identify system weaknesses and CAPEC to classify attack patterns, allowing standardized vulnerability reporting and assessment in robots and drones.



\subsubsection{\textbf{Authentication and Authorization Failures}}

\textbf{\textit{Existing Threats:}}
The Common Weakness Enumeration (CWE-287), Improper Authentication, defines this vulnerability as occurring when an actor claims to have a given identity, but the product does not prove or insufficiently proves that the claim is correct. Such a weakness can enable unauthorized entities to impersonate legitimate users or devices, potentially leading to system compromise and loss of control. One of its associated attacks, CAPEC-114, Authentication Abuse, describes a scenario in which an attacker obtains unauthorized access to an application, service, or device by exploiting flaws or weaknesses in the authentication process. According to \cite{koubaa2019micro}, authentication in sUASs ensures that each node can verify the origin of transmitted data, thereby confirming that messages are received from legitimate and trusted sources. The authentication of the unmanned system by the Ground Control Station (GCS) is therefore critical to guarantee that the GCS is communicating with an authorized drone rather than a malicious or spoofed one. Similarly, the authentication of the GCS itself is equally important, ensuring that the sUAS neither transmits sensitive data nor executes commands issued by a compromised or fake control station. Furthermore, as highlighted by \cite{park2023provably}, sUASs are especially vulnerable to drone capture attacks, in which adversaries physically seize and manipulate drones to extract cryptographic material or reprogram their internal control logic. These threats underscore the importance of implementing robust mutual authentication mechanisms between sUASs and GCSs to safeguard against both remote impersonation and physical compromise.


\textbf{\textit{Defense Mechanisms:}}
To mitigate the risks associated with improper authentication researchers have proposed several defense mechanisms aimed at strengthening the authentication process in sUASs. According to \cite{koubaa2019micro}, secure communication between sUASs and GCS should incorporate mutual authentication protocols to ensure that both entities can verify each other’s legitimacy before any data exchange occurs. This mutual verification process prevents unauthorized entities from injecting malicious commands or intercepting sensitive data. The authors also emphasize the importance of using cryptographic keys and certificates to validate identities, as well as incorporating encryption to protect the confidentiality and integrity of control and telemetry messages. In addition, Park \textit{et al.}~\cite{park2023provably} propose a provably secure mutual authentication and key agreement scheme that leverages Physical Unclonable Functions (PUFs) to generate unique hardware-based identities for each sUAS. This approach ensures that even if a drone is physically captured, its authentication credentials cannot be cloned or reused by an adversary. The integration of lightweight cryptographic operations and session key establishment in their framework significantly reduces the risk of impersonation and replay attacks while maintaining computational efficiency suitable for resource-constrained sUAS platforms. Overall, these defense mechanisms—mutual authentication, encryption, and PUF-based identity verification—collectively enhance the resilience of sUAS communication networks. They ensure that only authenticated and authorized entities participate in mission-critical communication, effectively mitigating the threats from improper authentication and drone capture attacks.


\textbf{\textit{Defense Gaps and Open Problems:}}
Despite advances in sUAS authentication, several gaps remain. Traditional mutual authentication schemes often rely on heavy cryptographic operations, which can strain the limited computational and energy resources of sUASs. Many current solutions also assume trusted communication channels or centralized control, making them vulnerable to spoofing, relay, and capture attacks. Zero Trust Architecture (ZTA)~\cite{stafford2020zero}, offers promising principles—such as continuous verification, least-privilege access, and identity-based security—that could enhance sUAS network protection. Its main advantages are improved resilience against insider threats and dynamic authentication across distributed nodes. However, implementing ZTA in sUAS systems presents challenges, including increased communication overhead, latency, and the need for constant identity validation, which may degrade real-time performance and power efficiency. In summary, while ZTA can strengthen authentication by removing implicit trust, adapting it to lightweight, latency-sensitive sUAS environments remains an open research challenge.


\subsubsection{\textbf{Encryption Absence or Weakness}}


\textbf{\textit{Existing Threats:}}
CAPEC-97 defines Cryptanalysis attack pattern as the process of identifying weaknesses in cryptographic algorithms and exploiting them to decrypt ciphertext without access to the secret key. While this definition is general, it is highly relevant to sUASs, which often rely on weak or absent encryption. For instance, \cite{koubaa2019micro} identifies critical encryption vulnerabilities in the MAVLink protocol, widely used for communication between sUASs and GCSs. Both MAVLink v1 and v2 transmit messages in plain text, providing neither confidentiality nor robust authentication. As a result, attackers can intercept, modify, or replay commands, enabling MitM and spoofing attacks. Although MAVLink v2 introduced an optional message-signing mechanism, it does not provide end-to-end encryption, and its adoption is limited by hardware and performance constraints. Further evidence of the risks posed by weak encryption comes from penetration testing conducted during a sUAS swarm competition sponsored by BAE Systems where \cite{alhawi2019finding} demonstrated that non-encrypted sUASs, such as the Tello and Parrot Bebop 2, could be compromised through malformed connection requests that overloaded the CPU buffer or by injecting forged control packets into the communication channel. The study also highlights the effectiveness of fuzzing techniques in identifying software vulnerabilities, emphasizing their value in improving sUAS system security.



\textbf{\textit{Defense Mechanisms:}}
Several studies have proposed hardware- and software-based security mechanisms to strengthen the MAVLink communication protocol used in sUAS systems. In \cite{koubaa2019micro, shoufan2015secure}, a lightweight hardware solution was reported to secure communication between the GCS and the sUAS using an FPGA-based cryptographic module. This module employs AES-CBC-MAC to encrypt and authenticate both command and payload data exchanged between the sUAS and the GCS. While this approach enhances data confidentiality and integrity, it also introduces performance overhead and increases power consumption due to the added hardware components. Similarly, a few researches describe a Raspberry Pi–based encrypted communication channel designed to maintain control of the sUAS in the event of an ongoing attack. Although this redundancy improves system resilience, it results in communication latency and higher CPU utilization, potentially affecting real-time sUAS performance~\cite{yoon2017security,koubaa2019micro}. From a software perspective, Hamza \textit{et al.}~\cite{hamza2024mavlink} highlights encryption-based defenses to protect the confidentiality of sUAS communications. The proposed framework, MAVSec, integrates encryption algorithms such as RC4, AES-CBC, AES-CTR, and ChaCha20 to prevent eavesdropping and data modification within MAVLink messages. Collectively, these defense mechanisms align with ZTA principles \cite{stafford2020zero}, which emphasize continuous authentication, full encryption of communication channels, and isolation of potentially compromised nodes. 



 \textbf{\textit{Defense Gaps and Open Problems:}}
 Despite the progress in securing MAVLink communications through hardware and software-based defenses, several limitations and research gaps remain.  For instance, Gambo \textit{et al.}~\cite{gambo2025zero} acknowledge that while adding encryption improves security, it also introduces significant challenges and trade-offs. First, encryption in resource-constrained sUAS platforms often results in higher processing overhead, which can degrade real-time control responsiveness and increase latency. Second, encrypted communications require additional power consumption and possibly heavier or more capable hardware—factors that are critical for sUASs with limited weight and energy budgets. Third, implementing full end-to-end cryptography and key management in existing MAVLink deployments faces compatibility and integration issues, because many systems were designed for minimal overhead, not robust security. Finally, even when message-signing (as in MAVLink 2.0) is available, the uptake remains limited, meaning the actual benefit of encryption remains unverified in many operational sUASs.


\subsubsection{\textbf{Remote ID Exploitation}}


\textbf{\textit{Existing Threats:}} 
sUASs generally operate under specific communication protocols to ensure standardized data exchange. In particular, within wireless communication systems, Remote ID is implemented to identify and distinguish the source of data signals, providing information about the vehicle and its operator. According to the FAA Remote ID regulation~\cite{FAA_FAR89}, this identification system provides real-time flight information about operating sUASs. The transmitted Remote ID data typically include the sUAS’s identity (e.g., serial number or session ID), the geographic coordinates (latitude, longitude, and altitude) of both the aircraft and its control station, as well as the sUAS’s velocity, timestamp, and emergency status information. By continuously sharing such Remote ID information, the system ensures a clear identification of each vehicle in operation and facilitates the acquisition of essential operational details necessary for regulatory and administrative oversight.

The Remote ID, which is essential for the operation of sUASs, contains critical information and can be exposed to various risks under different circumstances. The potential exposure of Remote ID can be broadly classified into two categories: the direct leakage of the Remote ID itself and the subsequent exploitation of leaked information to conduct attacks on the sUAS system. In fact, in 2022, a controversy arose when the DJI drone company experienced a data breach that exposed over 80,000 drone IDs~\cite{news2022}. This type of incident represents a risk of Remote ID exposure, which may stem from inadequate data privacy management. However, it can also occur through other means, such as signal sniffing or network hacking, whereby an attacker could identify and disclose the Remote ID of an sUAS. Even if an attacker does not obtain the Remote ID itself, intercepting the overall data packet may enable a replay attack, allowing the attacker to impersonate a legitimate user and gain unauthorized access by exploiting the victim’s communication channel. Furthermore, once a Remote ID is compromised, attackers can use the stolen information in various ways. They might use signal spoofing to cause deliberate obfuscation~\cite{srimoungchanh2024assessing}. This creates a vulnerability, letting attackers pose as other users, transmit false information, or access communication networks. The attacker may also track the drone using the compromised ID~\cite{kerns2014unmanned}. More critically, they can track the user or operator who controls the drone~\cite{radovs2024recent}. In addition, when individuals maliciously misuse the Remote ID system, they may broadcast falsified information about non-existing sUASs, creating misleading signals that can disrupt normal operations and interfere with legitimate flight activities.



\textbf{\textit{Defense Mechanisms:}}
To protect the Remote ID of sUASs from exposure to such attacks, various defense methods have been investigated. Representative approaches include strengthening security by adding an authentication step to the communication protocol, and enhancing the privacy of the Remote ID through improved encryption techniques. Therefore, recent studies have explored various defense mechanisms to prevent threats arising from the hijacking or unauthorized use of Remote ID systems.

First, several methods have been proposed to enable privacy-preserving transmission of Remote ID messages during broadcasting. These approaches primarily aim to enhance the security of broadcast information through message encryption. For instance, the Anonymous Direct Authentication and Remote Identification (A$^{2}$RID)~\cite{wisse20232} protocol introduces a novel scheme that can be applied to heterogeneous commercial drones. In this system, each sUAS broadcasts anonymous Remote ID messages. The protocol employs a group signature mechanism that utilizes group public key and private key, allowing authorized entities to decrypt messages accordingly. This group-based public key structure enables additional authentication, restricting access to those who do not possess the appropriate group keys, thereby securing Remote ID messages and their associated data. Similarly, the selective authenticated pilot location disclosure (SNELL)~\cite{tedeschi2024selective} protocol is designed to allow sUASs to transmit Remote ID messages containing encrypted information about the pilot’s location. Within this protocol, the Ciphertext-Policy Attribute-Based Encryption (CP-ABE~\cite{bethencourt2007ciphertext}) and Schnorr-based signature schemes \cite{seurin2012exact} are employed to ensure that only authorized recipients specified by the sUAS can decrypt the actual location data. These cryptographic enhancements strengthen the overall security of Remote ID transmission protocols by mitigating potential vulnerabilities.

Beyond privacy-guaranteed communication protocol designs, other studies have applied Differential Privacy techniques to preserve location privacy by generating Remote IDs that include perturbed rather than actual coordinates~\cite{brighente2024obfuscated,brighente2022hide}. Moreover, blockchain-based frameworks \cite{wu2021blockchain} have been proposed to replace actual IDs with generated ID addresses. Such methods maintain privacy for both identity and location data while enabling decentralized and distributed ID management. Collectively, these approaches aim to prevent external attackers from persistently tracking specific sUASs based on the information contained in Remote ID broadcasts.



\textbf{\textit{Defense Gaps and Open Problems:}}
In accordance with sUAS-related regulations currently being implemented in various countries, including the United States (FAA), Europe (EASA), and Japan (JCAB), each sUAS is required to broadcast a Remote ID containing diverse operational information~\cite{tedeschi2023privacy}. While embedding extensive information within a Remote ID and sharing it transparently allows neighboring sUAS operators to utilize it more effectively and enhances overall operational management; however, as previously mentioned, inherent vulnerabilities in the Remote ID system and the potential for identity spoofing pose serious security and privacy concerns. Consequently, a fundamental trade-off exists between transparency and privacy regarding the extent of information that should be disclosed through Remote ID broadcasts.

Although multiple approaches have been studied to mitigate vulnerabilities associated with Remote ID, there remains a significant gap between current research-level solutions and technologies that are ready for real-world deployment. To be viable for deployment, such technologies must demonstrate robustness not only against a single vulnerability but simultaneously across multiple threats. However, existing methods, even those incorporating obfuscation mechanisms to preserve privacy, remain susceptible to direct telemetry broadcast tracking and are ineffective against replay and linkage attacks inferred through temporal and spatial inference. Due to these persistent limitations, current regulations have yet to incorporate standardized, privacy-preserving Remote ID solutions that meet regulatory acceptance. In other words, despite the numerous vulnerabilities introduced by the broadcast nature of Remote IDs, existing regulations do not yet provide comprehensive measures to completely mitigate these risks.

Therefore, for future directions, it is essential to develop Remote ID broadcast regulations that adequately balance both privacy and transparency. To achieve this, a comprehensive privacy-preserving Remote ID protocol addressing the previously discussed limitations is required. In this context, several technical advancements are necessary, including adaptive privacy schemes that generate Remote IDs based on environmental and situational contexts, secure aggregation mechanisms among multiple USSs, and communication frameworks designed according to region- or geolocation-specific Remote ID policies.

\subsubsection{\textbf{Cloud-Based UAS Service Suppliers Exploitation (Third-Party Data Attacks)}}


\textbf{\textit{Existing Threats:}}
The UTM ecosystem is highly dependent upon cloud-hosted UAS service suppliers (USS) or third-party information services to obtain mission-critical data such as weather and wind fields, terrain and obstacle databases, and traffic and telemetry feeds from cooperative participants \cite{kopardekar2016utm,faautm2020}. These services are typically accessed through application programming interfaces (APIs) that follow protocols such as Representational State Transfer (REST) over Hypertext Transfer Protocol Secure (HTTPS), Message Queuing Telemetry Transport (MQTT), or other proprietary data feeds. This data is then utilized by the communication and planning layers of the CNS architecture (see Fig.~1). This tight coupling creates an attack surface in which a compromised or malicious data provider, a tampered feed, or even a stale or delayed update can propagate directly into the decision making process such as trajectory generation, geofencing, and separation-assurance. Common attack vectors include API credential theft, MitM attacks on unverified Transport Layer Security (TLS) connections, Domain Name System (DNS) spoofing to redirect clients, and data poisoning in weather or traffic datasets \cite{sedjelmaci2017uav,khan2017iotsec}. Similar data-poisoning and integrity attacks have been observed in cloud-connected aviation and autonomous ground transportation systems, where adversaries compromise or spoof third-party information services, such as weather, navigation, or map databases, to mislead the decision-making process and create unsafe behaviors \cite{manesh2017adsb,mccallie2011security,zhao2021securecacc,sharma2022v2x}.


\textbf{\textit{Defense Mechanisms:}} Mitigating these threats requires end-to-end authentication, integrity checking, and redundancy in cloud data utilization. Authenticated APIs should use cryptographically signed requests, mutual TLS (mTLS) for server–client verification, and periodic API key rotation to reduce impersonation risks \cite{rescorla2018tls,hardt2012oauth}. Data source verification mechanisms, such as validating digital signatures and timestamps, help ensure that cloud-provided information originates from trusted sources. Redundancy and cross-validation are also important. UTM clients should compare data from multiple providers, reconcile discrepancies using onboard models, or default to authoritative sources like the FAA System Wide Information Management (SWIM) infrastructure \cite{swim2019faa}. Timeout and fail-safe behaviors should ensure that if cloud data becomes unavailable or unverifiable, the system reverts to conservative flight envelopes or return-to-base routines. Replay and spoofing detection can be achieved by embedding timestamps or sequence numbers in API responses.


\textbf{\textit{Defense Gaps and Open Problems:}}
Despite these defenses, the current UTM concept of operation lack standardized frameworks for verifying the trustworthiness of cloud-sourced data \cite{faautm2020}. Many sUAS implementations assume that any HTTPS endpoint is secure, creating overreliance on single providers or unauthenticated public APIs. In addition, verifying third-party data in real time is computationally challenging for small aerial vehicles with limited onboard resources. Future work should focus on secure-by-design data management pipelines, where each provider issues cryptographically signed and timestamped datasets, and federated verification services capable of detecting anomalies or revoking compromised sources. Research into distributed trust registries, lightweight cryptographic attestation methods, and consensus-based validation across multiple cloud providers could form the basis for a trustworthy data ecosystem for UTM.

\subsection{\textit{Navigation Vulnerabilities and Defense Mechanisms}}

This section focuses on vulnerabilities that compromise the navigation subsystem of sUASs, particularly those relying on the GPS, INS, and related localization methods. Attacks in this category primarily aim to mislead, degrade, or deny positioning and velocity information used by the autopilot or flight controller. Detailed explanations of the navigation vulnerabilities, associated defense mechanism, and the research gaps as well as open problems are discussed hereunder: 

\subsubsection{\textbf{GPS Spoofing}}

\textbf{\textit{Existing Threats:}}  
GPS spoofing is the deliberate fabrication of satellite-navigation signals intended to mislead GPS receivers. Typically, a spoofer transmits counterfeit signals on the same frequencies used by GPS so the receiver locks onto them and reports a false position.  The effectiveness of an attack depends on the spoofer's sophistication: crude or abrupt falsifications are easy to detect, whereas a sophisticated attacker will first mimic authentic satellite signals and then slowly manipulate them, causing the receiver to drift off its nominal path without obvious warning \cite{PsiakiHumphreys2016,meng2022survey}.

GPS spoofing can create serious safety and security risks for aviation. By falsifying navigation fixes, an attacker can steer air vehicles off planned routes, such as into hostile territory where they could be captured or into restricted airspace. For example, media and agency reports in 2015-2016 raised concerns that U.S. Customs and Border Protection (CBP) drones operating near the U.S.–Mexico border experienced GPS spoofing that interfered with their mission \cite{Tucker2015DHS}. Civilian GPS receivers are particularly vulnerable because they are usually unencrypted and operate at a single frequency, lacking cryptographic authentication or multi-frequency verification capabilities \cite{PsiakiHumphreys2016,meng2022survey}.

Researchers study GPS spoofing using testbeds ranging from hobbyist software-defined radio workflows (e.g., GPS-SDR-SIM \cite{Ebinuma2015} combined with RTL-SDR/USRP hardware) to commercial GNSS simulators \cite{PsiakiHumphreys2016}. Table~\ref{tab:gps_spoofing_comparison} summarizes the key characteristics, tools, and consequences of different spoofing attack types.

\begin{table}[!t]
\centering
\caption{Comparison of GPS Spoofing Attack Types}
\renewcommand{\arraystretch}{1.2}
\setlength{\tabcolsep}{3.5pt}
\begin{tabular}{p{1.5cm} p{5cm} p{4.5cm} p{4.5cm}}
\toprule
\makecell[l]{\textbf{Spoofing} \\ \textbf{Type}} & \textbf{Key Characteristics} & \textbf{Typical Tools / Implementations} & \textbf{Operational Consequences} \\
\midrule
\vspace{5pt} 
\makecell[l]{\rotatebox{90}{\parbox[r]{1cm}{\centering\textbf{Coarse} \\ \textbf{(Overt)}}}}&
\begin{itemize}
    \item Abrupt signal power increase
    \item High-power takeover of authentic GPS
    \item Easily detected by receiver
\end{itemize} 
&
\begin{itemize}
    \item Basic SDR transmitters
    \item GPS-SDR-SIM with RTL-SDR
    \item Low-cost signal generators
\end{itemize} &
\begin{itemize}
    \item Sudden position jumps
    \item Loss of satellite lock
    \item Obvious navigation failure
\end{itemize} \\
\midrule
\vspace{0pt}
\makecell[l]{\rotatebox{90}{\parbox[r]{2cm}{\centering\textbf{Sophisticated}\\\textbf{(Stealthy)}}}} &
\begin{itemize}[leftmargin=*, nosep]
    \item Gradual, seamless signal takeover
    \item Closely mimics authentic satellite timing
    \item Enables “carry-off” manipulation
\end{itemize} &
\begin{itemize}[leftmargin=*, nosep]
    \item High-end GNSS simulators
    \item USRP-based spoofing platforms
    \item Multi-antenna phased systems
\end{itemize} &
\begin{itemize}[leftmargin=*, nosep]
    \item Gradual trajectory deviation
    \item Return-to-home malfunction
    \item Undetected mission deviation
    \item Potential airspace violation
\end{itemize} \\
\bottomrule
\end{tabular}
\label{tab:gps_spoofing_comparison}
\end{table}

\textbf{\textit{Defense Mechanisms:}}  
Recent research proposes several complementary strategies to detect and prevent GPS spoofing. Signal-level methods monitor Doppler shifts, carrier-to-noise ratios, and other anomalies that reveal inconsistencies between genuine and counterfeit signals \cite{manesh2019detection}. Authentication-based defenses such as Navigation Message Authentication \cite{fernandez2016navigation} and signal watermarking \cite{esswein2021gps} add cryptographic integrity to future GNSS designs, while multi-sensor fusion with inertial or visual odometry enables real-time cross-verification of navigation data \cite{meng2022survey}. Data-driven approaches further enhance detection by training machine learning (ML) models to recognize abnormal signal patterns \cite{dang20233d}. These techniques vary in complexity and computational cost: cryptographic methods provide strong guarantees with low runtime overhead  but require infrastructure upgrades, whereas ML-based methods are  computationally demanding. On the other hand, simple signal and sensor-fusion checks offer lightweight protection suitable for sUAS \cite{PsiakiHumphreys2016, meng2022survey}, though less effective against stealthy spoofing. A summary is provided in Table~\ref{tab:gps_defense_mechanisms}.

\begin{table}[!t]
\centering
\caption{GPS Spoofing Defense Mechanisms}
\renewcommand{\arraystretch}{1.15}
\setlength{\tabcolsep}{5pt}
\begin{tabular}{%
    p{2.8cm}  
    p{3.6cm}  
    p{2.2cm}  
    p{2.6cm}  
    p{3.4cm}  
}
\toprule
\textbf{Method} & \textbf{Requirements} & \textbf{Overhead} & \textbf{Spoofing Type} & \textbf{Trade-offs} \\
\midrule
\vspace{1pt}
Signal anomaly   \quad detection &
\begin{itemize}[leftmargin=*, nosep]
    \item GPS receiver
    \item C/N$_0$, SNR metrics
    \item Doppler monitoring
\end{itemize} &
\vspace{1pt}
Low &
\vspace{1pt}
Coarse or overt attacks &
\begin{itemize}[leftmargin=*, nosep]
    \item Simple implementation
    \item Ineffective against stealthy attacks
\end{itemize} \\
\midrule
\vspace{1pt}
Cryptographic \quad \quad authentication &
\begin{itemize}[leftmargin=*, nosep]
    \item Authenticated GNSS signals
    \item Crypto-capable receiver
\end{itemize} &
\vspace{1pt}
Low &
\vspace{1pt}
All spoofing types &
\begin{itemize}[leftmargin=*, nosep]
    \item Infrastructure-dependent
    \item Limited civilian availability
\end{itemize} \\
\midrule
\vspace{1pt}
Sensor fusion &
\begin{itemize}[leftmargin=*, nosep]
    \item IMU or INS unit
    \item Camera or LiDAR
    \item Visual odometry
\end{itemize} &
\vspace{1pt}
Moderate &
\vspace{1pt}
Coarse to sophisticated attacks &
\begin{itemize}[leftmargin=*, nosep]
    \item INS drift over time
    \item Requires environmental features
\end{itemize} \\
\midrule
\vspace{1pt}
ML methods &
\begin{itemize}[leftmargin=*, nosep]
    \item GPS receiver data
    \item Training datasets
    \item On-board compute
\end{itemize} &
\vspace{1pt}
Moderate to high &
\vspace{1pt}
Sophisticated or stealthy &
\begin{itemize}[leftmargin=*, nosep]
    \item Needs diverse training data
    \item Computational burden
\end{itemize} \\
\midrule
\vspace{1pt}
Signal watermarking &
\begin{itemize}[leftmargin=*, nosep]
    \item Infrastructure-level modifications
    \item Compatible receiver
\end{itemize} &
\vspace{1pt}
Low &
\vspace{1pt}
All spoofing types &
\begin{itemize}[leftmargin=*, nosep]
    \item Not yet deployed
    \item Requires global standardization
\end{itemize} \\
\bottomrule
\end{tabular}
\label{tab:gps_defense_mechanisms}
\end{table}

\textbf{\textit{Defense Gaps and Open Problems:}}  
Despite significant progress, several obstacles hinder practical spoofing resilience for sUAS. Authenticated GNSS signals remain largely unavailable to civilian operators, and cryptographic authentication is deployed only on military systems, leaving consumer drones vulnerable. Low-cost sUAS platforms typically lack the sensor redundancy required for robust multi-sensor fusion, restricting them to simple signal-anomaly checks that sophisticated spoofers can evade.

Stealthy spoofing attacks, in particular, pose a difficult detection problem. They mimic natural GPS drift or slow trajectory adjustments, so adversaries can remain undetected by threshold-based monitors. Machine-learning approaches show promise against such stealthy spoofing but face significant deployment challenges, including the scarcity of representative training datasets that capture diverse spoofing scenarios, computational constraints on resource-limited platforms, and the difficulty of validating classifier performance in real-world flight conditions \cite{alhoraibi2024detection}. Future research should prioritize lightweight cryptographic authentication schemes tailored for consumer drones, collaborative spoofing detection through swarm-based consensus algorithms that leverage multiple platforms’ observations, and adaptive flight-control policies that dynamically adjust autonomy levels or mission parameters when spoofing is suspected.

\subsubsection{\textbf{GPS Jamming}}

\textbf{\textit{Existing Threats:}}  
GPS jamming is the deliberate transmission of high-power radio-frequency interference on GNSS frequency bands to deny positioning services. Unlike spoofing, which carefully manipulates signals to mislead receivers into reporting false positions, jamming simply overwhelms legitimate satellite signals with noise or narrowband interference, preventing receivers from acquiring or tracking any signal. Because GPS signals arrive at Earth's surface with extremely low power (having travelled over thousands of kilometres from orbit) even a modest ground-based jammer transmitting on the L1 (1575.42 MHz), L2 (1227.60 MHz) or L5 (1176.45 MHz) civilian bands can deny service across a wide area. The consequences for sUAS are immediate and severe: loss of position fixes triggers autonomous flight-controllers to enter fail-safe modes, typically commanding the aircraft to hover in place, attempt return-to-home (which may fail without navigation), or execute emergency-landing procedures. Most commercial sUAS either rely exclusively on GPS for navigation, or lack high-quality IMUs or alternative navigation sensors capable of sustaining extended flight during GNSS outages, making jamming effectively deny mission capability and potentially cause loss of control or crashes \cite{khan2021gps}.

Real-world incidents highlight the prevalence and accessibility of GPS jamming. Jamming has been widely documented during NATO military exercises \cite{ODwyer2018NATO}, in Eastern European conflict zones (particularly near Ukraine and the Black Sea region) \cite{RFERL2023BlackSea}, and around sensitive facilities where operators deploy jamming to counter unauthorised drone activity. Commercially-available GPS jammers are inexpensive and readily accessible through online retailers, further lowering the barrier to deploying such attacks. The ease of jamming GPS signals combined with the widespread reliance of sUAS on GPS-only navigation creates a significant operational vulnerability for both civilian and military drone operations.

\textbf{\textit{Defense Mechanisms:}}  
Traditional anti-jamming techniques for GPS systems are similar to those for anti-spoofing. They include spread-spectrum methods such as frequency-hopping, which rapidly switches transmission frequencies to evade narrow-band jammers, and directional antennas with null-steering capabilities that spatially reject interference by adaptively placing nulls towards jamming sources \cite{10965638}. Advanced GNSS receivers employ adaptive filtering and signal-processing techniques to distinguish legitimate signals from interference. Multi-constellation GNSS receivers improve resilience by providing redundant signals across different frequency bands, complicating the jammer’s task~\cite{10140066, 10139969}. However, sufficiently powerful broadband jammers can still deny all constellations simultaneously.

For sUAS, practical defenses rely heavily on sensor redundancy. IMUs enable short-term inertial dead-reckoning when GPS is unavailable \cite{khan2021gps}, though accumulated drift limits their effectiveness to seconds or minutes depending on IMU quality. Visual odometry and vision-based simultaneous localization and mapping offer GPS-independent navigation by tracking features in camera imagery, as demonstrated by platforms like Skydio autonomous drones \cite{kendall2017end}. Barometric altimeters and magnetometers provide altitude and heading references that, when fused with IMU data, extend autonomous operation during GNSS-outages. However, vision-based methods require sufficient lighting and environmental texture, impose significant computational burdens, and may be unsuitable for high-speed or featureless flight regimes. This can limit their applicability across diverse operational scenarios.

\textbf{\textit{Defense Gaps and Open Problems:}}  
Despite these available techniques, significant obstacles limit their deployment on low-cost sUAS platforms. Advanced methods such as controlled-reception-pattern antennas, adaptive filtering, and encrypted signal-processing impose prohibitive SWaP and cost constraints, restricting them to military and high-end commercial systems. Most commercial off-the-shelf drones lack real-time switching architectures between GNSS and alternative navigation modes, instead relying on rigid fail-safe behaviours (hover in place or return-to-home) that activate when GPS is lost. These default responses may be unsafe in contested airspace where hovering exposes the platform to further attack, and return-to-home itself requires GPS availability. Future research should prioritize affordable hybrid navigation architectures that gracefully degrade and seamlessly transition between GPS, inertial, and vision-based modalities as reliability changes. Collaborative localization strategies, where sUAS swarms share relative-positioning information to maintain formation during partial GNSS denial, offer a promising avenue for resilience without expensive onboard hardware on each platform.

\subsubsection{\textbf{Inertial and Sensor-Based Attacks}}


\textbf{\textit{Existing Threats:}}
As crucial parts of the navigation component in sUASs, Inertial Navigation Systems (INSs) and Inertial Measurement Units (IMUs) are vulnerable to sophisticated spoofing attacks that exploit the physical characteristics of MEMS sensors. Acoustic injection attacks target resonance frequencies of accelerometers and gyroscopes (typically 15-25 kHz), causing sensors to register phantom accelerations that do not correspond to actual vehicle motion~\cite{trippel2017walnut}. Vibrational attacks similarly exploit mechanical resonances, while electromagnetic interference can induce false signals in sensor readout circuitry~\cite{kim2012cyber}. These disturbances inject false accelerations or angular-rate biases that accumulate over time, severely degrading navigation accuracy, particularly during GNSS outages when the flight controller relies solely on inertial measurements from the navigation component~\cite{rodday2016exploring}.
The vulnerability is amplified in tightly coupled GNSS/INS architectures within the navigation component. Sophisticated adversaries can transmit falsified GNSS signals through the vehicle-to-vehicle (V2V) communication channel synchronized with inertial sensor manipulation, where spoofed GNSS position masks inertial anomalies and corrupts the navigation filter~\cite{kerns2014unmanned}. This coordinated approach defeats traditional fault detection mechanisms that rely on GNSS/INS cross-validation, enabling trajectory hijacking that compromises the flight controller for landing zone redirection or controlled crashes. Habler \textit{et al.}~\cite{habler2023assessing} emphasized the importance of GNSS/INS integration for navigation integrity while highlighting vulnerabilities in loosely coupled architectures. Trippel \textit{et al.}~\cite{trippel2017walnut} experimentally demonstrated that acoustic signals can alter MEMS output by exploitation of resonances using readily available audio equipment. Tanıl \textit{et al.}~\cite{tanil2018ins} proposed integrity monitoring within GNSS/INS fusion focusing on innovation-based fault detection. Ai-Ping~\cite{hu2011camera} studied spoof propagation through coupled systems in sUAS and UTM environments, emphasizing residual analysis and cross-verification protocols that could leverage Camera \& LiDAR data.

\textbf{\textit{Defense Mechanisms:}}
To address vulnerabilities in navigational components and coordinated GNSS/INS spoofing attacks, researchers have developed multilayered defense strategies. Recent research emphasizes anomaly detection, integrity monitoring, and AI-assisted fusion approaches. To detect false accelerations and angular rate biases caused by acoustic injection and vibrational attacks on MEMS sensors, Tanl \textit{et al.}~\cite{tanil2018ins} introduced residual-based algorithms that monitor the innovation sequence, the difference between the predicted and measured sensor values, to detect anomaly. Their approach applies statistical hypothesis testing to innovation patterns, triggering alerts when residuals exceed expected bounds. Addressing coordinated GNSS/INS spoofing where falsified signals mask inertial anomalies, Ai-Ping~\cite{hu2011camera} emphasized cross-verification between multiple independent navigation sources (GNSS, inertial, visual odometry, barometric altimetry) to isolate corrupted inputs. By comparing outputs from multiple modalities, systems can identify outliers and dynamically adjust sensor weights based on relative trustworthiness.
To counter coordinated attacks that synchronize falsified GNSS signals with inertial sensor manipulation, Gurgu \textit{et al.}~\cite{gurgu2022vision} advanced AI-assisted spoof detection using deep learning for multi-sensor residual analysis. Their framework applies recurrent neural networks trained on benign and adversarial data to identify subtle inconsistencies across visual, inertial, and GNSS streams that evade traditional threshold-based detectors. The models recognize complex spoof signatures including gradual drift injection and coordinated multi-sensor attacks, enabling adaptive responses such as sensor reweighting or autonomous return-to-home procedures. These studies signal a transition from rule-based monitoring to data-driven adaptive integrity frameworks for sUAS and UTM infrastructures.

\textbf{\textit{Defense Gaps and Open Problems:}}
Existing defenses fail under real sUAS flight conditions, where vibration, temperature changes, and electromagnetic interference obscure spoofing signatures or cause false alarms in UTM operations~\cite{rodday2016exploring,kim2012cyber}. Real-world hybrid attacks that coordinate on GNSS, IMU, and other sensors further degrade reliability, as redundancy checks collapse when all inputs appear consistent~\cite{kerns2014unmanned,wang2023survey}. The computational overhead of deep learning approaches poses challenges for resource-constrained sUAS platforms~\cite{habler2023assessing}, while the lack of standardized adversarial datasets for inertial sensors hinders comparative evaluation~\cite{wang2023survey}. Future efforts should focus on physics-informed adaptive AI frameworks for multisensor fusion that integrate GNSS, IMU, visual odometry, and LiDAR data to improve the integrity of navigation components under adversarial noise~\cite{gurgu2022vision,raissi2019physics}. Robust state estimation, realistic testing protocols that incorporate environmental stressors~\cite{rodday2016exploring}, and flight validation on sUAS and UTM platforms are essential to ensure certification and operational trust~\cite{faa2025aam}.

\subsection{\textit{Perception / Vision-based Vulnerabilities}}

This section examines vulnerabilities that target camera- and vision-based perception subsystems of sUAS's autonomy stack. Focus on attacks that manipulate visual inputs (digital or physical) to induce incorrect perception outputs (detection, tracking, depth/pose estimation) and the downstream effects on navigation and control. In the following, different types of vision-based cyber-security threats with the associated defense mechanisms and research gaps in the defense strategies are discussed with detailed explanations:

\subsubsection{\textbf{Visual Sensor Spoofing}}
\textbf{\textit{Existing Threats:}}
Modern sUAS operations are critically dependent on visual perception: as Figure~\ref{fig:suas_architecture} illustrates, Camera \& LiDAR sensors supply the navigation component and flight controller with essential visual and depth data for real-time decision making. Visual sensor spoofing involves injecting falsified or manipulated visual data into these autonomous perception pipelines to mislead navigation and control systems. These attacks manifest in two primary forms: \textit{digital spoofing}--tampering with image streams during Video Transmission, inserting synthetic frames, or manipulating pixel data, and \textit{ physical spoofing}-- deploying adversarial patches, projected light patterns, reflective decoys, or carefully designed physical objects that exploit model vulnerabilities. Such attacks directly compromise critical vision pipelines including object detection, obstacle tracking, depth estimation, and semantic segmentation, leading to erroneous scene interpretation within both the sUAS and broader UTM coordination frameworks.

Du \textit{et al.}~\cite{du2022physical} demonstrated physical adversarial attacks on aerial object detectors, showing that subtle optical perturbations degrade sUAS perception under varying illumination and altitude. Khan \textit{et al.}~\cite{khan2024visual} revealed that real-time visual spoofing can insert or remove objects from camera feeds, producing false detections or missed obstacles without direct access to internal systems. These vulnerabilities result in catastrophic control outcomes: the Flight Controller receives corrupted perception data leading to incorrect landing zone identification, collision with undetected obstacles, and complete mission failure, while the GCS + sUAS Operator remains unaware due to manipulated Video Transmission feeds. In UTM environments where multiple sUAS operate in shared airspace, visual spoofing poses additional risks to deconfliction and cooperative navigation.

\begin{figure}[H]
\centering
\includegraphics[width=0.6\textwidth]{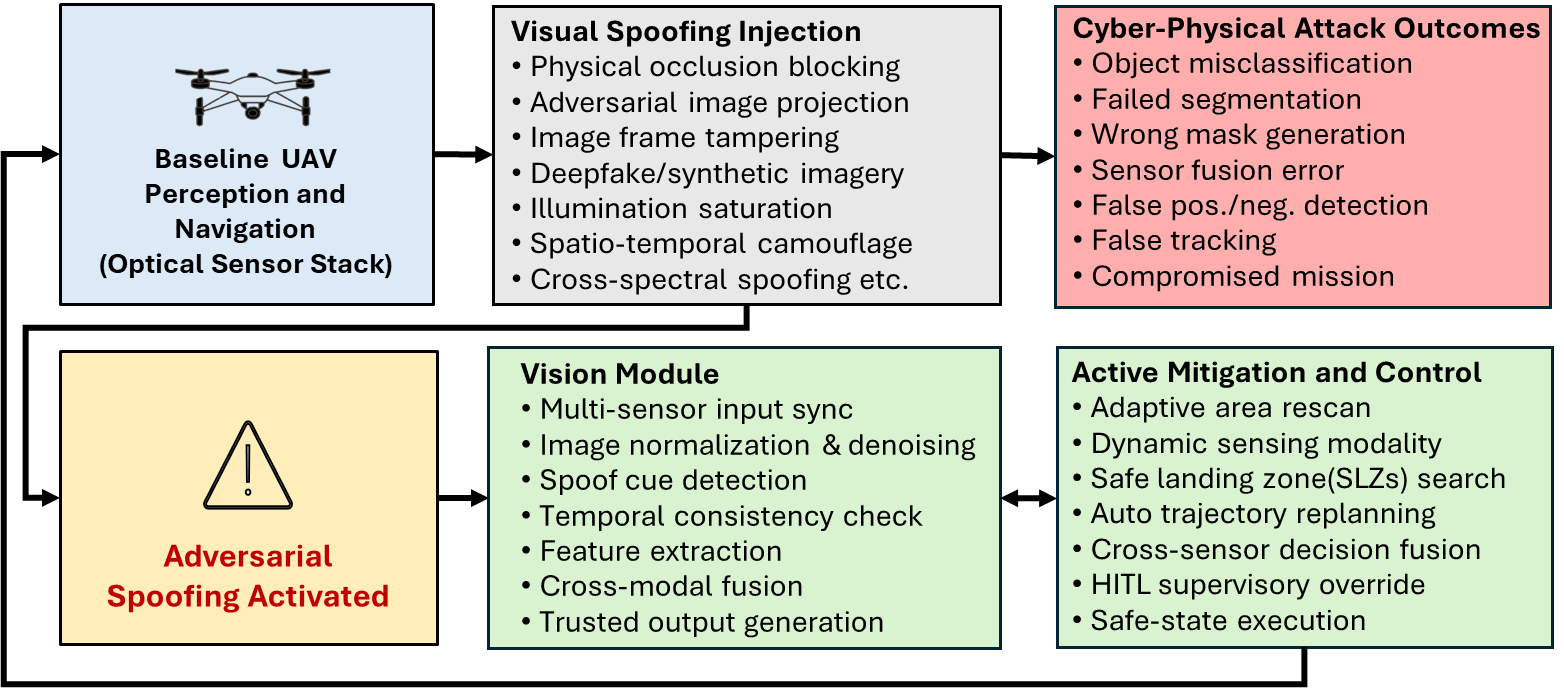}
\caption{UAS spoofing detection pipeline comparing clean baseline perception with adversarial activation and multi-sensor mitigation.}
\label{fig:detection_pipeline_tsu}
\end{figure}

\textbf{\textit{Defense Mechanisms:}}
To protect the Camera \& LiDAR subsystem from visual spoofing threats, defenses operate at multiple levels within the perception pipeline. Specifically targeting pixel-level perturbations and adversarial patches that degrade object detection and obstacle tracking, input filtering applies preprocessing such as Gaussian blurring, median filtering, and JPEG compression to suppress high-frequency adversarial perturbations before model inference~\cite{xu2017feature}. Addressing physical patch attacks and optical perturbations demonstrated by Du \textit{et al.} adversarial training incorporates attack examples during training to improve model robustness~\cite{madry2017towards}, although computational costs remain prohibitive for resource-constrained sUAS platforms. To detect digital manipulation and synthetic frame injection during video transmission, as shown by Khan \textit{et al.}'s real-time spoofing attacks, signal-level defenses leverage frequency-domain analysis using Fourier or wavelet transforms to detect spectral anomalies indicative of spoofing~\cite{xu2017feature}. Li \textit{et al.}~\cite{li2023semantic} proposed Semantic-SAM, a segmentation framework that enhances spoof resilience through semantic-level verification that detects inconsistencies in object classification that could lead to incorrect identification of Safe landing zones (SLZs). Canh \textit{et al.}~\cite{canh2024semantic} integrated semantic segmentation with sparse mapping for vision restoration in degraded conditions.

Physical verification through Camera–LiDAR fusion within the sensing subsystem enforces geometric consistency checks and rejects visually corrupted data before they reach the Flight Controller~\cite{xu2023real,hatch2021obstacle}. Defending against temporal attacks and frame manipulation in Video Transmission streams, Abady \textit{et al.}~\cite{lydia2022manipulation} exposed risks from synthetic deepfake imagery bypassing traditional detectors, motivating multimodal defenses. Spatial and temporal consistency checks, such as optical flow validation and frame correlation analysis, further reject inconsistent or suddenly altered visual input that could cause collision with undetected obstacles~\cite{muller2022physical,man2023person}. Multimodal fusion that leverages optical-LiDAR cross-validation within the Camera \& LiDAR subsystem provides geometric consistency checks to reject visually corrupted data and maintain scene integrity under adversarial conditions before propagating to the navigation component~\cite{xu2023real,yu2024physense}. Table~\ref{tab:defense_compact} summarizes the defense strategies already explored, their targeted attacks, and computational costs.

\begin{table}[H]
\caption{Comparison of visual spoofing defense strategies.}
\label{tab:defense_compact}
\centering
\small
\begin{tabular}{p{0.20\textwidth} p{0.24\textwidth} p{0.36\textwidth} p{0.10\textwidth}}
\hline
\textbf{Defense Type} & \textbf{Targeted Attack} & \textbf{Techniques} & \textbf{Cost} \\
\hline
Input Filtering & Pixel/projection noise & Blurring, denoising, compression~\cite{xu2017feature} & Low \\
Adversarial Training & Patch/perturbation attacks & Robust model training, augmentation~\cite{madry2017towards} & High \\
Signal Filtering & Frequency anomalies & Fourier/wavelet analysis~\cite{xu2017feature} & Low \\
Physical Verification & Cross-sensor spoofing & LiDAR–camera fusion, depth checks~\cite{xu2023real} & Medium \\
Temporal Consistency & Frame manipulation & Optical flow, correlation analysis~\cite{muller2022physical} & Medium \\
\hline
\end{tabular}
\end{table}

\textbf{\textit{Defense Gaps and Open Problems:}}
Despite progress in protecting the Camera \& LiDAR subsystem, visual spoofing defenses face significant real-world limitations. Environmental variations - dynamic lighting, shadows, fog, rain, lens flare - often mimic adversarial artifacts targeted by input filtering and signal-level defenses, causing false positives and reducing detection reliability against optical perturbations demonstrated by Du \textit{et al.}~\cite{du2022physical,man2023person}. Most defenses, including adversarial training and semantic segmentation approaches, are validated on static benchmark datasets (ImageNet, COCO) and controlled laboratory conditions, limiting generalization to real-time spoofing scenarios demonstrated by Khan \textit{et al.} across diverse camera types, sensor configurations, and operational environments~\cite{muller2024vogues,yu2024physense}. Critically, few techniques have been tested on actual sUAS hardware where the computational overhead of adversarial training and deep learning-based defenses clashes with vibration-induced motion blur, limited power budgets, and real-time processing requirements imposed by the Flight Controller~\cite{xu2023real,hatch2021obstacle}. The trade-off between detection accuracy and computational efficiency remains unresolved for edge deployment on resource-constrained sUAS platforms~\cite{habler2023assessing}, while adversarial robustness often degrades clean-sample performance~\cite{madry2017towards}.

Future research should prioritize developing lightweight spatio-temporal defenses that extend current temporal consistency checks to exploit motion patterns and multi-frame consistency with minimal computational overhead~\cite{muller2022physical,man2023person}. Adaptive multi-modal fusion architectures that dynamically adjust trust across Camera, LiDAR, and Navigation (INS, IMU) streams beyond static Camera-LiDAR fusion are essential for detecting coordinated visual and inertial spoofing in scalable AAM deployment~\cite{yu2024physense}. Establishing standardized adversarial sUAS benchmarks that incorporate real environmental dynamics, sensor noise profiles, and physical attack scenarios, including physical patches and synthetic frame injection threats, will enable a rigorous evaluation~\cite{wang2023survey,habler2023assessing}. Integrating online learning mechanisms for contextual adaptation will be critical for sustaining perception integrity within the Video Transmission and Command \& Control pathways against evolving spoofing threats in operational sUAS and UTM systems~\cite{rangi2021learning,khazraei2022learning}.

\subsubsection{\textbf{Target Misclassification / Vision Hijacking}}

\textbf{\textit{Existing Threats:}} Adversarial inputs cause sUAS detectors to mislabel objects or trackers to persistently lock onto wrong targets \cite{Chen2020OneShotDualAttention,Tian2021AdversarialAttacksUAVIoT}. Digital attacks exploit gradient sensitivities through imperceptible pixel-level perturbations that induce class confusion \cite{Shrestha2023RobustAdvPatchUAV}. Tracker poisoning introduces a single adversarial frame, causing cumulative drift \cite{Chen2020OneShotDualAttention}, while physical adversarial textures achieve continuous tracking failure \cite{Wiyatno2019PhysicalTexturesVOT}, both causing long-time target loss. Consequences include incorrect avoidance maneuvers, geofence violations, payload misdelivery, and unsafe landing approaches when adversarial markers corrupt landing-pad recognition. Standard evaluation metrics include mean Average Precision for detection degradation, Multi-Object Tracking Accuracy and Precision for tracking correctness \cite{Bernardin2008CLEARMOT}, identity switches, and recovery time measuring reacquisition latency after hijacking.

\textbf{\textit{Defense Mechanisms:}} Model-level defenses employ adversarial training on synthesized perturbations and heterogeneous ensemble fusion combining convolutional neural networks (CNN) and vision transformer backbones to diversify features and reduce transferability \cite{Lu2023UAVEnsembleRobustness}. Input-level defenses apply spatial smoothing and frequency-domain filtering to sanitize suspicious regions \cite{Shrestha2023RobustAdvPatchUAV}. Temporal consistency checks validate trajectories against kinematic priors while multi-frame re-detection resets corrupted tracker states when confidence drops \cite{Chen2020OneShotDualAttention}. Runtime anomaly detectors monitor activation statistics to flag adversarial frames without model modification, enabling lightweight deployment on resource-constrained platforms. Domain randomization and adversarial data augmentation during training enhance robustness, with sUAS-specific adaptations addressing camera jitter and altitude-dependent scale variation.

\textbf{\textit{Defense Gaps and Open Problems:}} The perception–control coupling problem remains critically underexplored. Current work narrowly focuses on detection metrics rather than downstream safety impact, failing to model how visual errors propagate through autopilot logic and mission planning to affect flight behavior. This gap is particularly acute for safety-critical applications where misclassification consequences extend beyond perception accuracy to physical system safety.

\subsubsection{\textbf{Stealthy Physical Perturbations}}

\textbf{\textit{Existing Threats:}} 
Physical-world modifications mislead sUAS vision systems while remaining inconspicuous to humans. Localized adversarial patches printed on surfaces achieve over 80\% attack success through Expectation-over-Transformation optimization accounting for altitude variation, viewing angles, and lighting changes \cite{Shrestha2023RobustAdvPatchUAV,Athalye2018SynthesizingRobustAdversarialExamples}. Environmental attacks manipulate ambient conditions through projected spotlights exploiting auto-exposure \cite{Zhou2022DoubleStar}, infrared lasers targeting sensor vulnerabilities \cite{Sato2024InvisibleReflectionsNDSS}, and adversarial shadows cast by obstacles \cite{Zhong2022ShadowsDangerousCVPR}. Recent diffusion models generate natural-looking patches matching environmental textures \cite{Lin2025DiffusionToConfusionICIP}. Physical attacks persist across missions without requiring system access and scale to entire fleets when deployed in operational areas, with motion blur and perspective distortion amplifying effectiveness.

\textbf{\textit{Defense Mechanisms:}} 
Temporal defenses include spatio-temporal consistency checks flagging kinematic violations \cite{Xu2024PhyScoutCCS} and multi-frame aggregation filtering transient perturbations. Multi-sensor fusion cross-validates visual detections with LiDAR, IMU, or depth data, though physically realized 3D adversarial objects can compromise all modalities simultaneously \cite{Cao2021InvisibleBothCameraLidarSP}. Contextual reasoning rejects physically implausible detections based on motion constraints. Model-agnostic preprocessing formulates patch mitigation as occlusion removal, achieving approximately 30\% attack success reduction without retraining \cite{Pathak2024ModelAgnosticDefenseUAVPatch}. Patch localization with generative inpainting detects high-frequency irregularities and reconstructs plausible backgrounds \cite{Pathak2024ModelAgnosticDefenseUAVPatch}. Hardware defenses include adaptive optical filters, while preprocessing applies Total Variation smoothing and Non-Printability Score filtering \cite{Shrestha2023RobustAdvPatchUAV}. Adversarial training with physical transformation augmentation improves robustness to real-world variations \cite{Lu2023UAVEnsembleRobustness,Shrestha2023RobustAdvPatchUAV}.

\textbf{\textit{Defense Gaps and Open Problems:}} 
Standardized sUAS adversarial benchmarks with systematic physical attack annotations are absent—current evaluations rely on clean datasets like VisDrone or COCO \cite{Zhu2021DTMDronesTPAMI,Lin2014COCO}. Physical validation remains limited, with most defenses tested exclusively in simulation or controlled labs unable to replicate outdoor conditions, material degradation, and deployment logistics encountered in field operations. Low-cost small unmanned aircraft systems relying solely on RGB cameras lack sensor redundancy for robust fusion, creating security disparity between military and commercial platforms. Multi-sensor integration and model ensembles exceed computational budgets of embedded processors under strict weight and power constraints. Emerging directions include runtime confidence monitoring and outlier detection, though maturity remains insufficient for deployment. Future benchmarks must incorporate realistic physical attack protocols and environmental diversity for reproducible evaluation.

\subsection{\textit{System and Software Vulnerabilities}}

This section examines vulnerabilities that arise at the system and software level of sUAS and UTM stacks. Focus on weaknesses that affect the integrity, authenticity, and availability of software and high-level mission data — including failures in mutual trust/attestation, tampering with mission plans and telemetry, and compromises during over-the-air data or software transfers. Detailed explanations of different categories of system and software vulnerabilities with defense mechanisms are as follows: 

\subsubsection{\textbf{Trust Validation (Mutual Attestation Failures)}}

\textbf{\textit{Existing Threats:}}
Trust validation is the process of verifying properties of an entity before taking privileged action based on interaction with it.
Examples within the Transport Layer Security~(TLS) commonly used on the internet
include validating the signature of a signed message for integrity
and validating the certificate chain of a server for authenticity.
Attestation is the process of providing evidence of identity and integrity to a verifier.
Mutual attestation is when two or more entities each attest and verify.
This goes beyond authentication in that evidence is provided not just for the identity of an entity,
but also for the integrity of the state of an entity.
A secure attestation scheme must be resilient to various known attack vectors
such as loading unsigned firmware,
bypassing secure boot,
replaying valid attestation tokens,
and compromising attestation infrastructure.
Vulnerabilities in trust validation can affect various system components.
A flight controller given invalid sensor data may make unsafe control actions.
A GCS given invalid sUAS telemetry data may issue unsafe commands to a sUAS.
A vulnerability in bootloader image validation may lead to loading software infected with malware.
A vulnerability in update client image validation may lead to loading firmware infected with malware.
A vulnerability in USS backend API validation may lead to publishing incorrect environment data.
Prior works have discussed sUAS-related firmware and attestation compromise  \cite{uav-sec-priv-survey, uavnet-sec-review}.
Tamper-proof logs and telemetry are necessary to detect trust violations.

\textbf{\textit{Defense Mechanisms:}}
There are several existing core defense mechanisms for trust validation.
Secure boot and measured boot mechanisms protect against bootloader vulnerabilities by verifying firmware signatures at startup.
Remote attestation can be used to prove software integrity to an external verifier.
Some platforms provide hardware-enforced isolation between execution contexts.
Such features can be used to create a Trusted Execution Environment~(TEE) to run sensitive code such as operations on cryptographic keys.
A prominent example for embedded systems is Arm TrustZone.
One way to improve trust validation in the supply chain is through the use of Software Bill of Materials~(SBOMs),
which are hierarchical inventories of software components and dependencies.
In the research program HACMS~\cite{HACMS},
formal methods were combined with model-based systems engineering to produce a secure experimental sUAS platform employing attestation mechanisms.
A recommended reporting format is Arm's Platform Security Architecture Attestation Token.

\textbf{\textit{Defense Gaps and Open Problems:}}
Due to hardware and resource constraints,
there is limited adoption of defense mechanisms in low-cost sUAS.
It may be difficult to manage attestation keys and verifiers at scale in future UTM-integrated ecosystems.
While a single sUAS operator may manage its own GCS and sUAS deployments,
if attestation is to occur between sUAS and USS managed by different operators,
then either entities may potentially need to be aware of valid keys for all other entities,
or some centralized PKI must be set up.
Furthermore,
different vendors or USSs may implement different protocols that are not natively interoperable,
indicating a need for either a compatibility service or a regulation mandating use of a standard attestation protocol.
Even if arbitrary pairs of prover and verifier entities are interoperable,
intermittent or lossy communication channels pose a threat to the reliability of attestation mechanisms.
Future research directions include lightweight, distributed attestation protocols and scalable integrity-checking infrastructures for sUAS fleets.

\subsubsection{\textbf{Data Tampering (Mission Plan Modification, Telemetry Changes)}}

\textbf{\textit{Existing Threats:}}
Data tampering is when an unauthorized modification is made.
In the context of sUAS and UTM,
high-value targets may include mission plans or telemetry for sUASs.
A data tampering attack may be perpetrated by a remote attacker through MitM or API key theft.
Such an attack could also be achieved through locally compromising a GCS.
Using data tampering capabilities,
an attacker may alter mission manifests,
inject fabricated telemetry data,
or manipulate databases and APIs.
Such behaviors may lead to sUASs deviating from planned flight paths,
masking of greater system failures,
or poisoning of data consumed by other systems such as AI.
MAVLink data tampering~\cite{MAVLink-sec-survey} is a well-known example that can lead to mission failure and loss of control.
Recommended detection signals include monitoring for inconsistent telemetry data and invalid waypoints.
Recommended reactive controls include safe-hold and revalidation.
When a security incident does occur,
it is important to have tamper-free evidence to submit for forensic analysis.
If logs are not secured,
then the data contained within them may be tampered with to prevent forensic analysis from identifying the cause of the incident.
It is important for timestamps of logged data to be accurate and for telemetry data to be cross-validated.

\textbf{\textit{Defense Mechanisms:}}
There are several existing core defense mechanisms for trust validation.
For message integrity verification,
there are cryptographic Message Authentication Codes~(MACs)
and digital signatures.
For log integrity verification,
there are hash chains, tamper-evident logging, and Merkle trees for mission records.
For cloud and USS access,
there are secure APIs and mutual authentication.
For redundant telemetry verification,
one can compare onboard, Remote ID, and USS feeds.
It is important to verify integrity in real-time as well,
using monitors and anomaly detectors on mission data streams.

\textbf{\textit{Defense Gaps and Open Problems:}}
Detecting subtle tampering in real-time is challenging because there are many compounding uncertainties in telemetry reporting.
Sensor errors, timing jitter, and control disturbances can all result in slight deviations from plans.
There is a lack of unified standards for telemetry logging and auditability across sUAS paltforms.
For external telemetry data,
there is an inherent trust boundary from an operator to a cloud service or USS service if the operator is not able to host their own instance of the service.
Future research directions include lightweight integrity attestation, secure data provenance tracking, and cross-layer telemetry validation frameworks.

\subsubsection{\textbf{Over-the-Air Updates \& Data Load Transfer Attacks}}

\textbf{\textit{Existing Threats:}}
Over-the-Air~(OTA) updates and data load operations are two types of data provisioning for remote hosts such as sUASs.
While data load operations provide non-executable data such as configuration files, map data, and missions,
OTA updates provide executable data such as firmware images, app updates, and security patches.
An adversary may attack these kinds of operations to tamper with firmware, install backdoors, persist data on the host, or exfiltrate data from the host.
Vectors for such attacks include unsigned udpates, rollbacks, compromised update servers, TLS bypass, and cellular MitM.
Effects of a successful attack may be full system compromise, removal of safety features, or permanent manipulation of mission behavior.

\textbf{\textit{Defense Mechanisms:}}
There are several existing core defense mechanisms for OTA updates and data load operations.
End-to-end code signing enforces digital signature checks for all updates,
ensuring the authenticity and integrity of the data received by the remote host.
Manifest validation ensures that the version, source, and compatibility checks of an update are cryptographically validated.
Rollback protection can be implemented to block the installation or reversion to outdated, vulnerable firmware.
Secure boot takes measurement of the firmware image at startup and prevents the bootloader from proceeding
if the firmware is unsigned, has an invalid signature, or has been modified and therefore does not match its signature.
When communicating the data to the remote host,
a protocol that ensures cryptographic freshness can be used to prevent replay attacks.
Such protocols include nonces or timestamps with payloads.
In many cases,
it is possible to use an existing secure channel solution such as TLS with mutual authentication.
SBOMs and attestation can even be used to provide more defense in-depth.
Defense mechanisms can be evaluated through simulated hijacking, rollback tests, recovery benchmarks, or OTA fuzzing.

\textbf{\textit{Defense Gaps and Open Problems:}}
Being SWaP-constrained systems,
many low-cost sUAS lack built-in support for these defense mechanisms
such as secure boot and verification hardware.
Even when supported,
management can become challenging,
such as preventing signing key compromise and rotating keys at fleet scale.
Performing more in-depth verification of delivered data on the remote host can consume significant time and power
if done offline when network access is intermittent during updates.
When updates include third-party blobs,
there is limited visibility unless complete SBOMs are provided.
Future research should look into scalable, distributed OTA verification,
lightweight secure update protocols, and secure update scheduling for energy-constrained sUAS.

\section{Conclusion}
\smallskip 

This survey presents a comprehensive, system-oriented review of cyber–security vulnerabilities and defense mechanisms for sUAS operating under the UTM framework. Unlike prior surveys that focus on high-level UAM security, our contribution lies in (i) mapping attacks directly to the communication, navigation, perception, and software subsystems that form the cyber–physical backbone of sUAS, (ii) synthesizing defense strategies according to their practicality for low-SWaP platforms, and (iii) identifying cross-layer interactions where failures can propagate across CNS, sensing, and software components. By organizing these findings into a unified taxonomy, we highlight key security gaps specific to UTM-integrated sUAS operations, including insufficient authentication and Remote ID privacy protections, limited resilience to GNSS spoofing and visual adversarial attacks, dependency on unverified cloud data, and the lack of secure and scalable OTA, attestation, and telemetry-integrity frameworks. Across all subsystems, a recurring research gap emerges: most existing defenses remain isolated, computationally expensive, or validated only in controlled environments. Practical deployment on resource-constrained sUAS requires lightweight, interoperable security solutions that account for environmental variability, multi-sensor inconsistency, and evolving adversarial capabilities.
Future directions include developing standardized adversarial testbeds for navigation and perception, lightweight cryptography and attestation tailored for small UASs, cooperative jamming and spoofing detection, robust vision/perception models, privacy-preserving yet regulation-compliant Remote ID protocols, and secure-by-design data pipelines for cloud-dependent UTM services. Advancing these areas will be essential for achieving secured, scalable, and trustworthy low-altitude sUAS operations and the UTM framework.

\section*{Acknowledgments}
\smallskip 

This material is based upon work supported by the NASA Aeronautics Research Mission Directorate (ARMD) University Leadership Initiative (ULI) under cooperative agreement number 80NSSC24M0070. Any opinions, findings, and conclusions or recommendations expressed in this material are those of the author(s) and do not necessarily reflect the views of the National Aeronautics and Space Administration.

\bibliography{sample}

\end{document}